# Aggregation of the web performance of internal university units as a method of quantitative analysis of a university system: the case of Spain

Enrique Orduña-Malea

Department of Audiovisual Communication, Documentation and History of Art. Polytechnic University of Valencia (UPV). Camino de Vera s/n, Valencia, 46022 (Spain). E-mail: enorma@upv.es, Tel: **+34** 963879480

**Abstract:** The aggregation of web performance (page count and visibility) of internal university units could constitute a more precise indicator than the overall web performance of the universities and, therefore, be of use in the design of university web rankings. In order to test this hypothesis, a longitudinal analysis of the internal units of the Spanish university system was conducted over the course of 2010. For the 13,800 URLs identified, page count and visibility was calculated using the Yahoo! API. The internal values obtained were aggregated by university and compared with the values obtained from the analysis of the universities' general URLs. The results indicate that, although the correlations between general and internal values are high, internal performance is low in comparison to general performance, and that they give rise to different performance rankings. The conclusion is that the aggregation of unit performance is of limited use due to the low levels of internal development of the websites, and so its use is not recommended for the design of rankings. Despite this, the internal analysis enabled the detection of, amongst other things, a low correlation between page count and visibility due to the widespread use of subdirectories and problems accessing certain content.

## Introduction

Whichever research method is used to analyze university performance (bibliometric, cybermetric, economic, etc.), the raw data obtained, if taken on their own, are not sufficient, either to describe the object of study (the university), or to draw conclusions about a particular aspect of study (performance, the relationship between variables, underlying causes, and so forth). The results need to be compared and, in order to do this, it is essential for them to be appropriately visualized. Therefore, in order to assess, we must first compare. The data obtained are relative units which require reference values in order for their magnitudes to be understood. The analysis of just one university would be meaningless: the results obtained for one university must be compared with the corresponding data for other universities in order to contextualize the magnitudes in any quantitative analysis.

In this regard, rankings are tools which enable us to visualize any research results comparatively. This technique consists of making a list of items which are ordered on the basis of a non-trivial criterion. Each item is characterized by a set of attributes (which are assumed to be essential to any description of the entity concerned), which are then used to obtain a final value, the criterion used to order the entities involved (Orduña-Malea, 2012).

The simplicity, ease of comprehension and the impact for those who view such rankings have ensured that they have become one of the most widely used tools in the dissemination of the results of university analysis over the first decade of the 21st century (Usher & Savino, 2007; Delgado López-Cózar, 2012). However, despite the advantages of rankings, their creation poses problems when one seeks to incorporate characteristics reflecting the multidimensional and diverse nature of organizations (which universities are). This situation has led to the appearance of different approaches to the creation of university rankings.

This first of these approaches aims to integrate the different attributes reflecting the multidimensional and diverse nature of each ranked organization. Adherents to this school include the ARWU ranking of Shanghai,[1] the Ranking Web of Universities,[2] and the HEEACT ranking of Taiwan.[3]

A second approach is critical of the above and opts for the creation of an independent ranking for each attribute measured (known as multi-ranking), thereby avoiding the use of combined indicators and offering partial rankings for the characterization of the university as a whole. The Scimago Institutions Rankings (SIR),[4] Leiden University's bibliometric ranking[5] and the Japanese Daigaku ranking[6] exemplify this approach.

The third approach focuses on thematic rankings. These are a specialization of bibliometric rankings and, regardless of whether a single final score is used, or several (one for each indicator), they are characterized by their focus on a specific thematic area. Examples include the ISI Rankings of Spanish Universities according to Fields and Scientific Disciplines (currently known as I-UGR).[7]

The main characteristic of the fourth approach is that one can generate a personalized ranking: the users actively participate in the material visualization process, choosing the universities, criteria and weightings. The concept of ranking is also somewhat diluted, moving towards other forms of non-ordered visualization based on groupings or clusters. Examples of this are the German CHE university ranking[8] and the Taiwanese College Navigator.[9]

Finally, there is a fifth approach, the main feature of which is a change in the unit of analysis, so that instead of measuring universities, they measure and compare various university units, such as departments or research groups. The Tilburg University Economic Schools Research Ranking[10] and the Global Ranking of Political Science Departments (Hix, 2004) follow this method.

All of the aforementioned approaches present various methodological problems. Rankings with overall scores use combined indicators which, from a mathematical point of view, have a series of insurmountable conceptual problems (Glänzel & Debackere, 2009). The drawback of multi-rankings is that the resulting rankings are incomplete and partial, from which the overall effectiveness of the universities cannot be inferred; thematic rankings suffer from similar problems. Personalized rankings, despite their undeniable advantages and usefulness as information resources, are not true rankings formally or structurally (Orduña-Malea, 2011).

The rankings of university units merit special attention: firstly, precisely because they do not list universities, but certain institutions; and secondly, because they have the advantage of providing greater detail and precision, as they focus the analysis on the performance of specific entities. They are therefore effective in detecting centers of excellence (and also of weakness) in universities, which remain hidden when the analysis is of a general nature. For example, Neri and Rodgers (2006) created a ranking of economics departments of Australian universities and they found that productivity was highly skewed within the departments. Furthermore, the detection of lower-performing units enables the identification of institutions with different missions and objectives (not just scientific ones). This may assist in the determination of certain characteristics reflecting the diverse nature of the university, enabling in turn the overall performance of the university to be described with greater precision.

However, the creation of university rankings using an analysis of their internal units presents certain additional methodological problems (great difficulty in accessing certain information, problems with the aggregation of variables and units, etc.), as well as being more demanding in terms of effort and time. In this respect, the application of cybermetric techniques offers a series of advantages for the analysis of internal university units:

- If the university and its different constituent units have a website and the hierarchy existing between them is sufficiently clear in their corresponding URLs, then it becomes possible to measure the same set of indicators at different levels of analysis.

- Moreover, the dynamic character of the Internet, the functioning of search engines (the fundamental tools for the collection of cybermetric data) and the scales of these indicators (at higher levels of magnitude than the bibliometric indicators) contribute to making ranking an appropriate technique for the description of the relative impact of the different units studied (Aguillo, Granadino, Ortega, & Prieto, 2006).

However, the great majority of cybermetric analyses of university units focus more on the specific performance of these units (and on the behavior of diverse indicators at various levels), than on their use in the creation of university rankings. The most studied internal academic unit in the literature is the department: Thomas and Willet (2000) studied departments of librarianship and information science; Tang and Thelwall (2003) US history departments; Li, Thelwall, Musgrove, and Wilkinson (2003) British computer science departments; Tang and Thelwall (2004) psychology and chemistry departments; Li, Thelwall, Wilkinson, and Musgrove (2005a; 2005b) Physics, Chemistry and Biology departments from Australia, Canada and UK; and Ortega (2007) departments (and research groups) in Spain. Finally, of special interest is Li's doctoral thesis (Li, 2005) on website interlinking between departments.

Other university units studied are research groups (Thelwall, Li, Barjak, & Robinson, 2008), schools (Chu, He, & Thelwall, 2003), as well as lecturers' personal pages (Thelwall, & Harries 2004a; 2004b; Barjak, Li, & Thelwall, 2007), which are currently the object of research for part of the Acumen[11] project.

The Cybermetrics Lab (part of the Spanish National Research Council), which is responsible for various cybermetric rankings, has also experimented with the analysis of units, specifically repositories (Aguillo, Ortega, Fernandez, & Utrilla, 2010), hospitals (Utrilla, Fernandez, Ortega, & Aguillo, 2009), business schools and research centers, although these units do not necessarily belong to university institutions. The web ranking of universities published by this same group is calculated using the universities' general URL (Aguillo, Ortega, & Fernández, 2008).

There is, therefore, something lacking in the design of university web rankings based on the aggregation of the performance of their internal units – not just of one type of unit (such as departments or schools), but of all those parts which can be considered functional units of the universities.

This aggregation procedure becomes meaningful if one begins with the premise that a link to a particular functional unit (research group, department, school, university product or service, amongst others) is more precise (from the semantic point of view) than a link leading to a university's general URL (the meaning of which is much more ambiguous). Therefore, the total number of links to a university's functional units could provide a more precise vision of that university's overall visibility. This same reasoning can be applied to any other cybermetric indicator (page count or web audience).

**Aims and objectives**

The main purpose of this study was, therefore, to conduct a multi-level cybermetric analysis of the internal parts of a university system (that of Spain in this case) with the aim of determining whether this provides precise results (in terms of page count and web visibility) and, therefore, whether it can improve the design of university web rankings.

Complementary to this were the following specific objectives:

- To assess whether there is a very pronounced uneven distribution (whereby only a few units determine the general performance of the university) or, on the other hand, despite the existence of high-performing units, the general values for the university are not determined by the units studied.

- To determine whether the cybermetric indicators studied ($Rs$, $Rv$, $WIF$, $r$; explained in next section), and the relationships between them, behave in the same way at the internal and overall levels.

- To compare the universities' overall performance (based on an analysis of the general URL) with the aggregated performance (based on the total of the performances of each internal unit's URL) in order to estimate the degree of representativeness thereof.

In order to fulfill these objectives, two consecutive stages were necessary: the first, focusing on the study of the internal structure of universities (identification and determination of the units and a study of their characteristics), which had already been completed in Orduña-Malea (in press), and a second stage centering on the application of cybermetric indicators to the units identified, along with the aggregation and analysis of the results. It is this second stage which this study is concerned with.

**Methodology**

For the purposes of clarity, the process by which the sample data (units and URLs) were obtained is explained first of all; next, the process of cybermetric data collection; and finally, the process used for the statistical analysis of the obtained data. It should be noted that the process by which the sample data were obtained is identical to that followed in Orduña-Malea (in press). This is detailed again here in order to contextualize the data examined and facilitate understanding of the present study.

*Data collection*

Firstly, the universities to be studied were identified and then the internal units of each were determined.

*Universities*

In 2010, the Spanish university system consisted of 76 universities (public and private). The official list of universities (and corresponding URLs) can be obtained from the Spanish Ministry of Education[12] and the *Conferencia de Rectores de las Universidades Españolas* (CRUE, the Committee of Vice-Chancellors of Spanish Universities).[13]

Apart from the official URLs (those indicated in the official sources consulted), the existence of "alias" domains were found (those that have the same second-level domains, but a different top-level domain; for example: "ub.es" and "ub.edu"), as well as "alternatives" (those with different second-level domains; for example: "upcomillas.es" and "upco.es") at different universities. The decision was therefore made to compile all these URLs (by surfing the internet and using search engines), as other sites might link to them and they might house relevant information.

*University units*

The first step was to establish the internal structure of the universities according to their activities. In this case, the following activities were considered: teaching, research, knowledge transfer, services and administration. These activities demonstrate the multidimensional nature of universities. Apart from the three classic activities, two complementary activities were added (services and administration), and, while they are not the primary objectives or part of the mission of the institutions, they do enable the university to function correctly; they are also liable to generate a large quantity of online documentation.

Having established these fundamental activities, the different types or categories of unit associated with each activity were identified, as can be seen in Table 1. The types of unit were divided up according to their nature, into institutions (for example, departments or research groups) and products (for example, repositories or blog platforms, etc.).

***Insert Table 1 here***

Finally, 23 types of entity were considered (17 types of institution and 6 types of product). In the case of the institutions, only those whose creation was regulated and legislated for were chosen. With regard to the products (and after an initial stage of exploration) only those present in at least ten different universities were taken into consideration (at the time this research took place, 2010) and which were liable to generate large quantities of documentation.

The internal units were located by means of manually browsing within each website, checking the existence of each unit identified and that the unit had been correctly assigned to the appropriate type. The unit of analysis was established at the "website" level, so that only those URLs (and therefore their corresponding units) that met the following criteria were used:

- Each unit had to correspond to a subdirectory or subdomain within the general academic web domain. This eliminated from the analysis all of those units with domains external to the official university domain and all those that consisted of a single webpage (a single HTML or XML file).

- The URL which identified the website could not be dynamic (non-user-friendly), due to the problems that such URLs posed for the measurement process.

The URLs of all the units that met these criteria were compiled, including the different aliases and alternative domains found –as was the case at the general university level.

In the case of redirections between these domains, the following steps were followed:

- If a valid URL (A) redirected to another non-valid URL (B), the first was taken into account, as it can be linked to external sites, but the second URL was not.

- If a valid URL (A) redirected to another valid URL (B), both were taken into account.

- If a URL did not work or did not direct to any resource (a broken link), but it was syntactically valid, it was taken into account.

The search, gathering and standardization process (both of units and URLs) was initially carried out between January and March 2010. This process was updated on three further occasions (in June, September and December 2010), to gather new data and correct possible errors. The sample obtained was finally composed by 7,391 internal units (divided into 7,098 institutions and 293 products), which were represented by 13,800 URLs (13,417 corresponding to institutions and 383 to products). By way of illustration, the sample categorized according to the types of unit is showed in Table 2, specifying the number of items and their corresponding URLs (Orduña-Malea, in press).

*Insert Table 2 here*

*Measurement of the sample*

For each of the URLs obtained at the previous stage, two searches were performed: for page count or size ("site:url") and visibility ("linkfromdomain:url –site:url"), every three months over the course of 2010 (in the last weeks of March, June, September and December), except for the units corresponding to archives, which were only measured from the second data collection stage onwards. All of these searches were carried out using the Yahoo! API, by means of the LexiURL application (now called Webometric Analyst)[14], with all of the queries being placed into text files prior to use.

*Sample processing*

The data obtained from LexiURL was then entered into various spreadsheets for analysis. First of all, the raw data from all the URLs (regarding both size and visibility) were scaled from 0 to 100, by means of a quotient transformation process with the aim of then being able to work with the mean relative representativeness factor for page count or size (Rs), and the mean relative representativeness factor for visibility (Rv), indicators which enable the raw values obtained for each item (university unit) to be contextualized in relation to the total value obtained by all items of the system (the Spanish university system) for a given period of time (2010).

In order to obtain the Rs value, the total page count value obtained for all the URLs corresponding to Spanish universities over the course of a month (the accumulated size $v_{sn}$) was taken to be equal to 100; consequently, the value for each URL was calculated as a proportion of this. After normalizing the results, a page count percentage, proportional to the total for all the universities compiled via Yahoo! for every month (in this case, 4) in which measurements were taken, was obtained; this percentage is the "relative representativeness".

Later, the mean of the $v_{sn}$ values for each sampling month was calculated and a value between 0 and 100 was also obtained; this was called the "mean relative representativeness factor for page count or size (Rs)". This factor could be calculated for any group of the websites sampled and for any period of time. Moreover, by substituting the overall page count value for that of the number of links ($v_{vn}$), the Rv value could be obtained for the same set of URLs (Orduña-Malea, Serrano-Cobos, Ontalba-Ruipérez, & Lloret-Romero, 2010).

The Rs and Rv values were calculated both for the set of universities (141 URLs) and for the university units (13,794 out of 13,800)[15]. The WIF was also calculated (at both the university and unit levels each time data were collected), along with the annual interest rate, r (%). The latter value, obtained by means of the compound interest rate formula, was used as a substitute for the statistical range in order to determine the growth rate (both in page count and visibility) for the period studied. All of these indicators and the meaning of their variables are summarized in Table 3.

*Insert Table 3 here*

Finally, the aggregation of items was initiated, at two levels:

- At university level: the values (of page count and visibility) of the different URLs of the same university were aggregated, obtaining a single value for the page count and visibility of each university (uni_1 aggregation).

- At unit level:

  a) By type of unit: the values of all units belonging to the same type (including all variants) were aggregated, obtaining a single value for page count and visibility for each type (type aggregation).

  b) By university: the values of all units belonging to the same university (including all variants) were aggregated, obtaining a single value for page count and visibility for each university (uni_2 aggregation).

For the unit-level aggregations, the Rs, Rv, r (%) and WIF values were also calculated, along with the Spearman correlations between Rs and Rv. A principal component analysis (PCA) was also carried out in order to determine the position of the aggregated units (types and universities) in relation to the page count and visibility variables.

**Results**

This section is divided into two: units and unit aggregations (types and universities).

*Analysis of university units*

This section concerns the analysis of university performance (for page count and visibility), in terms of the internal units identified earlier.

*Measurement of size*

The Dialnet service, from the University of La Rioja (UR), is of major importance, accounting for 23.43% of total representativeness. Its influence may well be greater still, as the second-placed entity, AIPSE (*Análise e intervención psico-socioeducativa* 1, department of Analysis and psycho-socio-educational intervention 1), at the University of Vigo (UVI) shows a series of inconsistencies in the raw data which seem to point to an error or some temporary growth which was not subsequently maintained.

The importance of units can also be seen in the percentage of these entities relative to the university to which they belong, called Rs factor (uni). Apart from Dialnet (98.31% of the size of the UR), also of major significance are the library of the Complutense University of Madrid (UCM) (40.62%), the virtual campus of University of Extremadura (UNEX) (74.69%), the repository at the Autonomic University of Barcelona (UAB) (44.20%), the catalog of the University of Pablo Olavide (UPO) (68.94%), and the *Escuela Superior de Ciencias Experimentales y Tecnología* (School of Experimental Sciences and Technology) at the King Juan Carlos University (URJC) (62.55%), amongst other large units. In a complementary fashion, the Rs factor (type) enables page count to be identified according to entity type, with the most eye-catching values being those of Dialnet (87.84% of the size of all the catalogs), the UCM library (58.17% of all the libraries analyzed), UNEX's virtual campus (56.24%), and the *Centro* *de Documentación Europea* (European documentation center) at the University of Alicante (UA) (51.63%). A detailed analysis of the top 20 units in terms of the Rs value is available in Annex I (Table I.1), where this same representativeness factor is also shown for each university it belongs to and for the type of unit.

Otherwise, the accumulated page count of the 13,794 units analyzed grew significantly during the period studied, going from 7,579,127 in March 2010, to 10,638,205 in December 2010 (taking into account the fact that archive centers were incorporated at the second data sampling stage). These data indicate an expansion of the academic space taken up by what are assumed to be essential university units.

In order to assess the growth of the units in the most appropriate way, the annual interest rate, r (%), is used. The results indicate a pronounced growth of UNEX's virtual campus (3.03%), and the UPO library catalog (3.47%), as well as the slightly negative growth of the library's digital collection at the UCM (-0.21%) and the UA's blog platform (-0.13%). The annual interest rate for each of the top 20 units according to Rs value is also offered in Table I.1 (Annex I).

Distribution by type of unit:

Figure 1 provides a graphical representation of the distribution of positions according to type of unit, and for each unit type, the number of items in the top 50, 100, 200, 500 and 1,000. It is more detailed in Table I.2 (Annex I), in which on the right one can see the percentage of the total number of items that the number of items represents for that unit type.

Although 13,794 URLs were identified, only the results for the top 1,000 positions are shown, as after this point, the results are practically 0. One can also see how the departments, faculties and research groups are the units with most items in the top 1,000, due to the high total number of items for these categories. However, analysis of the percentages shows that the repositories and blog platforms are those which obtain the best relative position. If the analysis is restricted to the top 100 (a more elitist perspective), this behavior is yet more pronounced.

*Insert Figure 1 here*

Distribution by university:

The UCM is the university with the highest number of units in the top 1,000 ranking according to Rs (86 units), followed by the UA (64), the Polytechnic University of Catalonia (UPC) (58) and the Polytechnic University of Madrid (UPM) (45). The performance of these universities is very homogeneous in all parts of the ranking studied. However, there are other universities with a high number of units in the top 1,000, but with few in the upper reaches (top 100, 200 or 500). For example, in the lower half of the top 1,000, there is a high percentage of units from universities such as the University of Santiago de Compostela (USC) (going from 12 units in the top 500 to 41 in the 501-1000 range), the University of Basque Country (EHU) (going from 8 to 27), or the Ramon Llull University (URA) (going from 9 to 23), amongst others.

For this reason, greater attention should be paid to the rankings corresponding to the top 200 and the top 500, as they are neither excessively elitist (as is the case with the top 50 and top 100, which are liable to be affected by universities with few units of large size), nor excessively crude (the top 1,000 tends to highlight the number of units pertaining to each university, rather than their size).

In the lower reaches of the ranking, there are 12 universities, all of them private institutions, without a single unit in the top 1,000, whereas in the very top part of the ranking (the top 50), 25 universities have at least one unit: the UA has the most, with 7 units, followed by the UCM with 6 and the UPC with 5.

Further information about the distribution of units is offered in Table I.3 (available in Annex I), which shows the distribution of units according to the university they belong to in the top 50, 100, 200, 500 and 1,000, only for the universities with at least 10 units in the top 1,000 of the Rs ranking.

*Measurement of visibility*

As is the case with the page count data, Dialnet is the unit with the highest representativeness value, although it is much lower than the comparative page count value (Rv=5.35), and it is followed by the UA's *Departamento de Lenguajes y Sistemas Informáticos* (department of computer languages and systems, Rv=2.74).

With respect to the Rv values according to unit type, Dialnet accounts for, as a mean average, 56.58% of the representativeness of all the catalogs, IE Blogs (from IE University) 32.70% of the blog platforms, the Bosch i Gimpera Foundation at the University of Barcelona (UB) 54.22% of the OTRIs (Offices of Research Results Transfer), the IDEC at Pompeu Fabra University (UPF) 45% of the lifelong learning services, and the *Dipòsit Digital de Documents* (digital repository of documents, UAB) 34.07% of the repositories. These data show how many of the top units according to Rv have very high percentages in their respective categories.

With regard to representativeness (Rv) according to university[16], unsurprisingly, high values are also observed. If top 20 units according to general Rv are taken into account, only one of these, belonging to the University of Murcia (UM), has an Rv (uni) value lower than 10. Using mean values, Dialnet accounts for 97.61% of the UR's links, IE Blogs 76.08%, and the UA's department of computer languages and systems 48.09%, reflecting a very uneven distribution at these universities. The data also reveal some unexpected behavior a priori, such as the poor performance of the departments (only 2 in the top 20) and the research groups, whose first item appears in 30[th] place (*Grupo de Estructuras de Datos y Lingüística Computacional I*, Data structures and computational linguistics I research group, at the University of Las Palmas de Gran Canaria (ULPGC), with Rv=0.44.

A total of 5 units belonging to university libraries appear in the top 20, although the first of these (the Library of the UPC 5; "bibliotecnica.upc.es"), demonstrated a significant decrease over the course of the study period (r=-1.14%), countered by an increase on the part of the Library of the UPC 4 ("bibliotecnica.upc.edu"), with very rapid growth in December (r=2.74%), constituting a clear example of a change of policy regarding the use of multiple domains. Other irregular behavior (in the top 20) was detected in the case of the Bosch i Gimpera Foundation (with a disproportionate representativeness value in June), and the La Salle School 1 (high representativeness in September).

Finally, blog platforms and libraries appeared prominently (when compared with their page count), which points to the great ability to attract links on the part of entities of this type in the Spanish academic system.

The accumulated quantity of external links, just as with the page count measurements, also increased significantly, from 1,944,877 in March to 2,322,631 in December. Table II.1 (Annex II) expands this information by including the 20 university units with the highest mean representativeness factor (relative in this case to visibility, Rv), as well as the total values accumulated by the 141 URLs of the university system.

Distribution by unit type:

Distribution by unit type is shown graphically in Figure 2. A basic analysis reveals high performance by key institutions (faculties, groups, departments and schools). In any case, these results must be contextualized by the total number of units by type. A percentage analysis demonstrates the high performance of some products (catalogs, blogs and repositories) although the raw values were lower than expected. Also notable are the poor performance, in percentage terms, of the research groups (only 3.10% are in the top 1,000 places), and the good results obtained by the libraries (8 units appear in the top 50). All percentage data for each unit type is additionally available in Table II.2 (Annex II).

*Insert Figure 2 here*

Distribution by university:

In this case, universities from Catalonia predominate, occupying the first 4 positions. The UAB (61 units in the top 1,000) stands in the first position, followed by the International University of Catalonia (UIC) (52), UPC (51) and UB (48). Nonetheless, the value for the UIC is artificial, as most of its entities figure between top 500 and top 1,000, given the multiplicity of domains.

The behavior of the UCM (the university with the largest overall size) is of special interest: it has no unit in the top 20 for Rv (and only 2 in the top 50), despite having several units (such as the institutional library and repository) which are among the largest in page count (showed previously in Annex I, Table I.1).

The performance of the private universities is also poor. Of the 15 universities with no units in the top 1,000, 14 of these are private universities. At the very highest level (top 50), and as with the Rs ranking, 25 universities have at least 1 unit, with the UB leading the way with 6, followed by the UAB and the UPC with 4 each.

The complete distribution of units according to the university they belong to (in the top 50, 100, 200, 500 and 1,000) is available in Table II.3 (Annex II), only for those universities with at least 10 units in the top 1,000.

*Comparison of size and visibility data*

The correlation between the Rs and Rv values for the total number of units studied (13,794) is high (R=0.704). However, as was remarked earlier, only approximately 1,000 units achieved significant results (in the case of both page count and visibility), so that the many units with no results distort this value. In order to avoid this effect, the correlation was recalculated by taking into account only the top 1,000 units for Rs (the units of greatest size), obtaining a notably lower value (R=0.422). If Dialnet is also excluded from the analysis (due to the influence it may have), the resultant correlation is lower still (R=0.417).

Figure 3 shows the dispersion between the Rs and Rv values (1,000 units with a superior Rs, excluding Dialnet), revealing the weak correlation between the two variables, which is influenced by the extreme values for page count (department of computer languages and systems) and for visibility (department of analysis and psycho-socio-educational intervention 1) of some units.

*Insert Figure 3 here*

The observed weak correlation indicates that the units of greatest size are not those that are most linked to. This observation is confirmed by the low web impact factors of the units of greatest size, which also remain stable over time (for example, Dialnet has a WIF=0.05 both in March and December). Despite the fact that the WIF is of no use in assessing the performance of units at an individual level, it does serve to indicate, overall, an anomaly in the system visibility values. Thus, 41 units have a WIF over 500 (and 150 have a value over 100), which clearly reflects poor performance in terms of visibility.

For illustrative purposes, Table 4 includes the units with the greatest asymmetries between page count and visibility data (due to higher or lower visibility with respect to size). Additionally, the WIF trend for the 20 larger units is available in Table III.1 (Annex III).

*Insert Table 4 here*

*Analysis of the unit aggregations*

In this section the results of the aggregation of university units are presented, for both unit type and university.

*By unit type*

With regard to page count, the catalogs and departments take up almost half of the representativeness between them (26.53% and 20.92% respectively). The other units are considerably further behind (7.65% for the research groups and 7.04 for the schools). In any case, the performance of the departments and research groups can be attributed to the long tail generated by the large quantity of items for these units, and not so much to the existence of large research groups or departments.

Regarding visibility, in this case the relative representativeness is more evenly distributed. It is important to note that the top positions are occupied by institution-type units (departments, faculties, research groups and schools); with regard to products, blog platforms are particularly prominent.

In order to complement these results, the aggregated raw values for each type of entity (for both page count and visibility) are provided in Annex IV. The corresponding relative representativeness factors, the annual interest rate, r (%), and WIF values are included as well.

Figure 4 shows the results of the principal component analysis for the aggregated units. Despite the fact only two variables were considered (page count and visibility), the graph enables the distribution of the unit types to be visualized two-dimensionally, as well as their relationship with each component.

*Insert Figure 4 here*

The WIF values obtained are, in general, very low: in fact, there is no value above 1 in any category (the visibility values are higher than those for size). This indicator is less stable over time, when all of the data collection periods are taken into account, than those at the unit level. This can be attributed to the fact that the values involved are larger and, therefore, so are the variations in the data between the different data collection periods (particularly noteworthy in this respect are, for example, the values obtained for the business schools in March, the OTRIs in June and the research centers in September). However, the WIF values enable us to observe a difference in behavior between institution units and product units. The products produce a very significant level of performance in terms of page count, but this is not matched by the level of visibility they achieve, which explains the fact that their WIF values are particularly low (apart from those for blog platforms).

Lastly, the correlation between the Rs and Rv values is very high (R=0.857), which indicates that the aggregated values (categories instead of units) are favor the correlation between page count and visibility by attenuating the differences generated by specific units.

*Aggregated units by university*

The data demonstrate that the internal results account for relatively low percentages in comparison to the general ones. At the accumulated level (aggregating the values of the 76 universities), the accumulated internal page count (10,638,205) accounts for 51.16% of the general accumulated page count (20,794,776), whereas the accumulated internal visibility (2,322,631) accounts for only 22.25% of the accumulated figure at the general level (10,437,067). Despite this, the correlations observed between the data for general and internal page count (R=0.911) and for general and internal visibility (R=0.887) are very high.

This may indicate that the internal analysis may not introduce excessive differences, despite providing a low percentage of the general performance of the university. However, there are significant changes in the relative positions of the universities for relative representativeness (page count and visibility) between general and internal values. In fact, a mean difference of 6.84 positions was found for the Rs ranking, and 7.74 for the Rv ranking.

With regard to visibility data, it should be noted that, despite the fact they were low for both levels in comparison with the page count data, the internal percentage is especially low, which is coherent with the low levels for visibility and WIF of the non-aggregated units, seen previously. In fact, the WIF values for the universities at a general level are much higher (and more variable over time) than those obtained at the internal level (which is to be expected, given the greater quantity of links involved). This is especially significant for some universities, such as Madrid Open University (UDIMA) ($WIF_{int}$=8.00; $WIF_{gen}$=51.26), Catholic University of Valencia (UCV) ($WIF_{int}$=0.51; $WIF_{gen}$=10.34), although there are some exceptions, such as the University of Vic (UVIC) ($WIF_{int}$=4.60; $WIF_{gen}$=2.40).

In any case, these results are influenced both by the policies governing the creation of independent internal units –the International University of La Rioja (UNIR) is a clear case of a university which cannot be measured at an internal level as no units appropriate for study have been created, reflecting other characteristics of this institution- and by possible search engine anomalies.

In the latter respect, the page count measurements demonstrated some inconsistencies (internal page count greater than general page count) for the following universities: UR (101.94%), UA (105.54%), URJC (106.90%) and the University of Huelva (UHU) (182.61%), whereas for the measurement of visibility, inconsistent data were only found in the case of the UIC (137.08%). Of these problems, the only significant ones are those concerning the UHU (due to a change in the design of the website during the study period) and the UIC (due to the complexity of this university's information architecture, along with problems with the version in different languages and the use of multiple domains).

The correlation between the Rs and Rv values is very high, at both the general level (R=0.887) and the internal level (R=0.831). These data serve to confirm that the correlation between the two variables grows as the level of aggregation increases.

Finally, Figure 5 shows the results of the principal component analysis for the aggregated internal units, showing only those universities with a page count value above 50,000 units (39 universities). On the left, the same analysis can be seen for all 76 universities, showing the general position of the complete Spanish university system.

The data reflect the high representativeness of the UR (Rs=23.82) and the UCM (Rs=9.45) for page count, and of the UB (Rs=7.37), UPC (Rs=7.03) and UAB (Rs=6.08) for visibility, while the UA (Rs=7.99; Rv=5.70) and UPM (Rs=4.98; Rv=6.23) demonstrate more even levels of performance for both components.

*Insert Figure 5 here*

The predominance of universities from Catalonia for visibility is determined by the large quantity of links between the different alias URLs that these universities possess. For example, considering the last data set of the sample (December 2010), "uab.cat" receives 75,300 inlinks from "uab.es", "upc.es" receives 49,700 from "upc.edu", and "ub.edu" gets 34,700 from "ub.es". For that reason, the influence of the external links from alias URLs (at both the general and internal levels) is obvious, although the analysis of this is beyond the scope of this study, due to its complexity.

The complete results obtained for the aggregated units by university are shown in Annex V (Table V.1), including, for reasons of space, only those values corresponding to the December data collection. On the one hand, the general results, which were obtained by the aggregation of all the URLs (aliases and alternatives) obtained for each university, are shown. On the other hand, the internal results, which were obtained by aggregating all of the units pertaining to each university, are also displayed. It is thus possible to compare the performance of a given university in terms of its official URL with that corresponding to the sum of all internal functional units. Both for the general and internal results, the page count and visibility data (raw data, annual interest rate and the relative representativeness) are included and the WIF is also incorporated. The differences (in percentage terms) between the raw general results and the internal ones are also shown.

**Discussion**

Following the same order as the previous section, the results for the university units will be discussed first, followed by a discussion of those for the unit aggregations.

*University units*

*Page count Measurement*

The internal analysis indicates that a single unit (Dialnet at the UR) accounts for 23.43% of the representativeness for page count (Rs) for all the internal units of all the URLs of all the universities. Also prominent are the department of Analysis and psycho-socio-educational intervention at the UVI (Rs=5.50), the library at the UCM (3.82) and the virtual campus of the UNEX (Rs=2.80). These "superentities" account for large percentages of the whole university system, and even more so of the universities they belong to. In terms of scale, Dialnet makes up 98.31% of the Rs of the whole of the UR. In the case of the UCM, the library accounts for 40.62%, E-Prints Complutense 11.29% and the digital collection 10.22%. In other words, a very large part of the total page count is accounted for by just a few units, reflecting a very unequal distribution at this university. These results are particularly significant, as the page count distribution analysis shows that Yahoo! only detects values for approximately 1,000 units.

The important role with regard to page count of the product-type units (repositories, blogs, virtual campuses and, especially, catalogs) should also be noted. Despite the small number of items in these categories, they are large and growing.

Research groups stand out in a negative sense, despite being the unit type with the most items. Only 4 URLs for such groups appear in the top 50 URLs for Rs. The performance of research centers and institutions is also modest. The entity types with the greatest presence in the top 100 are: departments (20), schools (13), repositories (11) and faculties (10), while research groups have only 7 units, constituting a very low percentage given the large quantity of items for research groups (2,867).

At university level, the relative representativeness factor for the page count of the internal units reveals the poor performance of the private universities, practically all of which are at the bottom of the ranking. The best-placed private university is the University of Navarra (UNAV), in 31$^{st}$ place (Rs=0.63).

A more elitist analysis can be carried out by counting the number of university units that each university possesses (in the unit ranking for Rs) in the top 100. The data place the UCM, the UA and the UPM in the top places (with 8 entities each), followed by the UPC and the UAB (with 6 entities each).

*Measurement of visibility*

Dialnet is again the unit with the highest representativeness (Rv=5.35), followed by the department of computer languages and systems at the UA (Rv=2.74) although, as their Rv values indicate, the distribution is more even than that observed for size.

The presence of service and product institutions is much more pronounced than was the case for page count. Of the 20 units with the highest Rv, there are five libraries, 3 blog platforms and 1 repository, revealing the ability of these units to attract external links.

With respect to the distribution of units with the greatest representativeness, and taking the top 100 URLs into account, the departments perform best (with 16 units), followed by the research groups (with 14, but only 7 of these are in the top 50), and schools (12). In contrast, if we take a more elitist perspective by looking at the top 50, the positions change slightly: departments (9), libraries (8) and research groups (7). Catalogs (4) and blog platforms (4) are also prominent. On the negative side, the poor positioning of the business schools is significant (not one in the top 200 positions).

With regard to the positions achieved by the universities, those with a prominent presence in the top 100 include the UAB (9 units), the UB (7) and the UAB, the UPM and the University of Castilla-La Mancha (ULCM) (with 6 each). If one considers only the top 50, the UB (with 6) is the university with the most units, followed by the UAB and the UPC (with 4 each). Universities from Catalonia are therefore dominant, with strong performances also coming from the polytechnic universities.

*General*

The correlation between page count and visibility (calculated using the respective Rs and Rv representativeness factors) is very low (R=0.422) if only those units with at least a minimally significant performance are taken into account (1,000 units). This, along with the low WIF values, indicates a very low internal visibility when compared to the internal size. The phenomenon may be caused by problems accessing various university units (such as some virtual campuses and catalogs, which can only be viewed by authorized persons, with only the initial interface otherwise visible), limiting the likelihood of these being linked by external users. The syntax of the internal URLs may also be another limiting factor, of a subjective and objective type.

The subjective factor is that internal URLs tend to be long and complex, which negatively impact on the probability of them being used as links, while the objective factor is a technical issue affecting the ability to efficiently calculate the links to internal URLs, mainly owing to the use of dynamic non-user-friendly URLs and subdirectories (instead of subdomains).

Dynamic URLs were not included in the study sample and so did not impact upon the visibility measured. In contrast, subdirectories were included and they are a very significant presence: it is estimated that, for the same sample of units used in this study, out of the total number of subdomains and subdirectories, 81.44% are subdomains (Orduña-Malea, in press). This represents a methodological limitation with regard to the measurement of links at internal levels, as Yahoo! only calculates the links for the exact URL which is provided to it, and not for each of the files that makes up a particular directory (something which does not occur in the case of the subdomains). This means that the units studied may potentially receive more links than those which are shown by Yahoo!. However, a file by file analysis of the links is not feasible due to the time and cost involved, which means that the analysis of the unit's highest-level URL, even if it is a subdirectory, is the only manner in which to proceed. Although it is impossible to know how many links are lost thereby, there is a certain tendency to link to the homepage in any case (Thelwall, 2011), minimizing the effects of this somewhat.

*Aggregated units*

The results obtained for the aggregated units by type and then by university are discussed below.

*Unit type*

The catalogs and departments are the units with the greatest relative representativeness for page count (26.53% and 20.92% respectively), while the business schools and archives are the units with the lowest representativeness (0.10% and 0.07% respectively). In terms of visibility, the units with the best Rv are the departments (16.08%), the faculties (12.82%) and the research groups (12.07%), while the business schools, archives and the video platforms are those with the lowest values (all three with 0.23%).

Despite the good results in terms of page count for the product units (especially repositories, virtual campuses and catalogs), the analysis of the relative representativeness for visibility indicates that this type of unit did not achieve the results which one would expect, given their high performance in terms of page count. Only the blog platforms improve on their performance for visibility in comparison with that for page count. Thus, the catalogs and video platforms decrease in importance, particularly in the case of the repositories, which, despite having 6 units in the Rs top 50, has just one in the Rv ranking. This is corroborated by the low WIF values observed. This poor performance for visibility could be attributed to the existence of subdirectories, commented upon earlier. However, Orduña-Malea (in press) indicate a greater use of subdomains for product-type units (specifically, it is estimated that, of the 383 URLs referring to product units, 74.67% use a subdomain syntax). For this reason, the lack of links to units of this type is associated with access problems.

In both cases (page count and visibility), the effects of scale must be taken into account, i.e. the types of units with the greatest number of items are the most likely to achieve the highest Rs and Rv values, as many small quantities of data are added together: the effect of the long tail.

*Universities*

The university with the highest Rs is the UR (Rs=23.82%, due to the specific influence of Dialnet), followed by the UCM (Rs=9.45%) and the UA (Rs=7.99%). The University of Seville (US), which stands out for size in particular, something which was detected by Orduña-Malea et al (2009), falls down the ranking somewhat to $14^{th}$ position (Rs=1.45%). With regard to visibility, the Catalan and polytechnic universities predominate: UB (7.37%), UPC (7.03%), UPM (6.23%) and UAB (6.08%), followed by the UA (5.70%) and the UR (4.11%).

The values obtained for the universities from the general URLs and those obtained from the aggregation of the URLs of the internal units show that the latter account for a low percentage compared to the former, for page count (51.16%) and especially for visibility (22.25%). Despite this low representativeness of the internal units, the correlations between the general and the internal values are very high, for both page count (R=0.911) and visibility (R=0.887). However, the differences in the positions of the universities between the general and internal Rs and Rv rankings are large. This may be due to large number of universities with very low values, which may distort the calculated correlation values.

In addition, caution should be exercised due to certain limitations of the aggregation process, mainly concerning the existence of anomalies in the functioning of the search engines. Moreover, the links between different URLs of the same university (both at a general and internal level) and the effects of the widespread use of subdirectories must be studied in greater detail. In any case, these effects may have an impact on the results, but it is estimated that they will not substantially change them.

*General*

Finally, the correlation between the Rs and Rv values was calculated at three levels of analysis: by unit (R=0.422; N=1,000), university (R=0.831; N=76), and type of unit (R=0.857; N=23). These results indicate an increase in the correlation as the unit of aggregation gets larger. Besides the increase in the correlation between the variables of page count and visibility in the aggregations, an increase in the WIF was also observed (along with its variability between data collection periods). These results may indicate that the WIF provides more information when it is applied to internal rather than aggregated levels, and that it should never be used to assess units individually – rather, it should be examined at the system level to detect possible anomalies between the obtained page count and visibility.

**Conclusions**

The principal conclusions reached are detailed below.

*Uneven distribution and the existence of "superunits"*

At item level, an uneven distribution was found, both for page count and visibility (although in the latter case it was less pronounced), so that only approximately the first 1,000 URLs (out of a total of 13,794) achieve relative representativeness, with the rest appearing empty to Yahoo!. This issue points to the existence of many sites with little content or few links, and to the existence of a few sites with an abundance of both.

The high-performing units, or "superunits", possess very high relative representativeness percentages (for page count and visibility). Dialnet is particularly prominent ($Rs=23.43\%$; $Rv=5.64\%$). These entities skew the representation values (for type of entity and university), and so their effect should be appropriately weighted in university web rankings.

*Differences between the performances of institutions and products*

The units corresponding to products perform much better, taking scale into account, than the institutional units. Catalogs, repositories and virtual campus platforms achieved high page count values, taking into account the small number of associated items. Moreover, it should be remembered that units of this type tend to have a single item per university.

In contrast, the visibility data are particularly low for the product units, with only the blog platforms standing out. This can be attributed to access problems, due to the fact that these products often house material restricted to the university community. In the case of the video platforms, their low general values can partly be attributed to the fact that they are present at a still limited number of universities and also to the fact that they as yet possess little material.

The results obtained for the research groups and departments are very low. Although their relative representativeness values are high, this is an effect of the long tail generated by the uneven distribution, along with the high number of associated items (sum of many small values).

*Differences between the page count and visibility performances*

The internal level analysis indicated a very weak correlation between page count and visibility, which is corrected as units are aggregated into larger entities (types and universities), and this is mainly due to the low visibility values (detected at all levels of analysis). In the case of institutional units, this effect can be attributed mainly to the excessive use of subdirectories, whereas in the case of products, it can be attributed to problems accessing restricted content, and the as yet low number of associated items for the universities analyzed.

*Representativeness of the units*

The representativeness of the aggregated units in relation to the general values for the universities is very low. The total accumulated page count at the unit level accounts for just 51.23% of the total accumulated page count at the general level (December 2010). For visibility, this percentage is lower still, at 22.31%.

Leaving aside the visibility problems commented upon earlier, these results point to the existence of page count and external links assigned to

- the university as a whole, due to reasons related to document management at the academic websites, or

- entities not considered in this study, and therefore, not essential to a university.

Also, despite the high correlation between the general and internal values for page count (R=0.911) and visibility (R=0.887), the ranking for university performance based on unit performance provides different results to that of the general performance ranking. The reason for this is that the results largely depend on the internal structure of the universities' websites, the URL syntax, and use of multiple domains. In this sense, and given the results of this study, the online Spanish university system is not yet mature enough for an internal cybermetric analysis which can be used to design rankings. Such an analysis is, however, useful to determine this internal structure, as well as to describe the performance of university activities and to detect anomalies and inconsistencies.

In any case, this analysis should be replicated for other university systems (with different internal structures) in order to ascertain whether the results obtained may be generalized, something which is of particular interest for global web rankings.

Equally, with the incorporation of new entities, the advances being made by products (such as repositories and video and blog platforms), the creation of channels on social media sites, as well as the development of the internal structure of academic websites, further research will be necessary in order to gain greater knowledge about the web performance of internal university units.

*Private universities*

Finally, given that few units belonging to private Spanish universities were identified, and the poor performance of these (of 27 private universities, 12 were without a single URL amongst the top 1,000 for Rs, and 15 had none in the top 1,000 for Rv), the performance analysis at unit level for these universities is not considered to be representative, although it does point to a deficiency with regard to ensuring that the structure and functions of the universities are adequately reflected on the Internet.

**Endnotes**

1. Academic Ranking of World Universities. Retrieved September 3, 2012, from http://www.arwu.org

2. Ranking Web of Universities. Retrieved September 3, 2012, from http://www.webometrics.info

3. Performance Ranking of Scientific Papers for World Universities. Retrieved September 3, 2012, from http://ranking.heeact.edu.tw

4. The Scimago Institutions Rankings (SIR). Retrieved November 21, 2012, from http://www.scimagoir.com

5. Leiden Ranking – 2009. Retrieved September 3, 2012, from http://conference.cwts.nl/projects/leiden-ranking-2009

6. The Asahi Shimbun Company. Retrieved September 3, 2012, from http://www.asahi.com

7. Rankings I-UGR. Retrieved September 3, 2012, from http://www.rankinguniversidades.es

8. CHE Ranking. Retrieved September 3, 2012, from http://www.che-ranking.de

9. College Navigator in Taiwan. Retrieved September 3, 2012, from http://cnt.heeact.edu.tw

10. Tilburg University Top 100 of Economics Schools Research Ranking. Retrieved September 3, 2012, from https://econtop.uvt.nl

11. Acumen (Academic Careers Understood Through Measurement and Norms). Retrieved September 3, 2012, from http://research-acumen.eu

12. Ministry of Education, Culture and Sport - Spanish universities. Retrieved September 3, 2012, from http://www.educacion.gob.es/educacion/universidades/educacion-superior-universitaria/que-estudiar-donde/universidades-espanolas.html

13. Conference of Rectors of Spanish Universities. Retrieved September 3, 2012, from http://www.crue.org

14. LexiURL Searcher. Retrieved May 1, 2011, from http://lexiurl.wlv.ac.uk

15. Although a total of 13,800 units were identified, the internal analysis was applied only to 13,794, as 6 URLs were eliminated as they disappeared during the study period.

16. It should be recalled that the total number of links for the university (and in the same way, the total for size) is the total accumulated from the sum of all the internal units, and not the general total calculated from the university's URL.

**Figure legends**

FIG. 1. Graphical distribution of entities by category for the Rs value.

FIG. 2. Graphical distribution of units by type for the Rv value.

FIG. 3. Dispersion between Rs and Rv (n=1000).

FIG. 4. Principal component analysis (PCA) for the unit types.

FIG. 5. Principal component analysis for the aggregated units by university.